\documentclass[10pt]{llncs}
\usepackage{amsfonts}

\newcommand{\set}[1]{\left\{ #1\right\}}
\newcommand{\gilt}{:}
\newcommand{\sodass}{\,:\,}
\newcommand{\setGilt}[2]{\left\{ #1\sodass #2\right\}}

\newcommand{\realrange}[2]{\left[#1, #2\right]}

\newcommand{\unitrange}[2]{\realrange{0}{1}}

\newcommand{\Oh}[1]{\mathcal{O}\!\left( #1\right)}

\newcommand{\llabel}[1]{\label{\labelprefix:#1}}
\newcommand{\labelprefix}{} % later redefined using renewcommand

\newcommand{\discussionsize}{\small}

\newcommand{\frage}[1]{}

\newenvironment{code}{\noindent%\sf%
\begin{tabbing}%
\hspace{2em}\=\hspace{2em}\=\hspace{2em}\=\hspace{2em}\=\hspace{2em}\=%
\hspace{2em}\=\hspace{2em}\=\hspace{2em}\=\hspace{2em}\=\hspace{2em}\=%
\kill}{\end{tabbing}}

\newcommand{\labelcommand}{}
\newcommand{\captiontext}{}
\newsavebox{\codeparam}
\newcounter{lineNumber}
\newenvironment{disscodepos}[3]{%
\renewcommand{\labelcommand}{#2}%
\renewcommand{\captiontext}{#3}%
\sbox{\codeparam}{\parbox{\textwidth}{#3}}%
\begin{figure}[#1]\begin{center}\begin{code}\setcounter{lineNumber}{1}}{%
\end{code}\end{center}\caption{\llabel{\labelcommand}\captiontext}\end{figure}}

{\end{disscodepos}}

\newcommand{\Is}       {:=}

\newdimen\endofsize\endofsize=0.5em
\def\endofbeweis{~\quad\hglue\hsize minus\hsize
                 \hbox{\vrule height \endofsize width
\endofsize}\par}
\usepackage[utf8]{inputenc}
\usepackage{algorithmic}
\usepackage{algorithm}
\usepackage{wrapfig}
\usepackage{amsmath}
\usepackage{amssymb}
\usepackage{color}
\usepackage{latexsym}
\usepackage{makeidx}
\usepackage{array}
\usepackage{multicol}
\usepackage{numprint}
\usepackage{t1enc}
\usepackage{times}
\usepackage{graphicx}
\usepackage{url}
\usepackage{todonotes}
\npdecimalsign{.} % we want . not , in numbers

\definecolor{mygrey}{gray}{0.75}
\newcommand{\ie}{i.e.\ }
\newcommand{\etal}{et~al.\ }
\newcommand{\eg}{e.g.\ }

\def\MdR{\ensuremath{\mathbb{R}}}

\newcommand{\hmey}[1]{}
\newcommand{\sout}[1]{}

\newcommand{\mytitle}{Partitioning Complex Networks\\via Size-constrained Clustering}
\begin{document}
\title{\mytitle}
\author{Henning Meyerhenke, Peter Sanders and Christian Schulz\\ 
        \textit{Karlsruhe Institute of Technology (KIT)},
        \textit{Karlsruhe, Germany} \\
       \textit{Email: \{\protect\url{meyerhenke}, \protect\url{sanders}, \protect\url{christian.schulz}\}\protect\url{@kit.edu}} 
      }

\date{}

\institute{}

\maketitle
\begin{abstract}
The most commonly used method to tackle the graph partitioning problem in practice is the multilevel approach.
During a coarsening phase, a multilevel graph partitioning algorithm reduces the graph size by iteratively contracting nodes and edges until the graph is small enough to be partitioned by some other algorithm. A partition of the input graph is then constructed by successively transferring the solution to the next finer graph and applying a local search algorithm to improve the current solution.

In this paper, we describe a novel approach to partition graphs effectively especially if the networks have a highly irregular structure. 
More precisely, our algorithm provides graph coarsening by iteratively contracting size-constrained clusterings that are computed using a label propagation algorithm. The \emph{same} algorithm that provides the size-constrained clusterings can also be used during uncoarsening as a fast and simple local search algorithm. 

Depending on the algorithm's configuration, we are able to compute partitions of very high quality outperforming all competitors, or partitions that are comparable to the best competitor in terms of quality, hMetis, while being nearly an order of magnitude faster on average.
The fastest configuration partitions the largest graph available to us with 3.3 billion edges  using a single machine in about ten minutes while cutting less than half of the edges than  the fastest competitor, kMetis. 
\end{abstract}
\thispagestyle{empty}

\section{Introduction}

Graph partitioning (GP) is very important for processing very large graphs, \eg networks stemming from finite element methods, route
planning, social networks or web graphs.
Often the node set of such graphs needs to be partitioned (or clustered) such that there are few edges between the blocks (node subsets, parts).  In
particular, when you process a graph in parallel on $k$ PEs (processing
elements), you often want to partition the graph into $k$ blocks of (about) equal
size. Then each PE owns a roughly equally sized part of the graph.
In this paper we focus on a version of the problem that constrains the
maximum block size to $(1+\epsilon)$ times the average block size and tries to
minimize the total cut size, i.e., the number of edges that run between blocks.
Such edges are supposed to model the communication at block boundaries between the corresponding PEs.
It is well-known that there are more realistic (but more complicated) objective
functions involving also the block that is worst and the number of its
neighboring nodes~\cite{HendricksonK00}, but the cut size has been the predominant optimization criterion.
% since it correlates with  other formulations for a variety of graph classes. 
The GP problem is NP-complete \cite{Garey1974} and there is no approximation algorithm with a constant ratio factor for general graphs \cite{BuiJ92}. Therefore heuristic algorithms are used in practice.  

\begin{wrapfigure}{r}{0.475\textwidth}
%\begin{figure}[t!]
\vspace*{-.75cm}
\begin{center}
\includegraphics[width=0.475\textwidth]{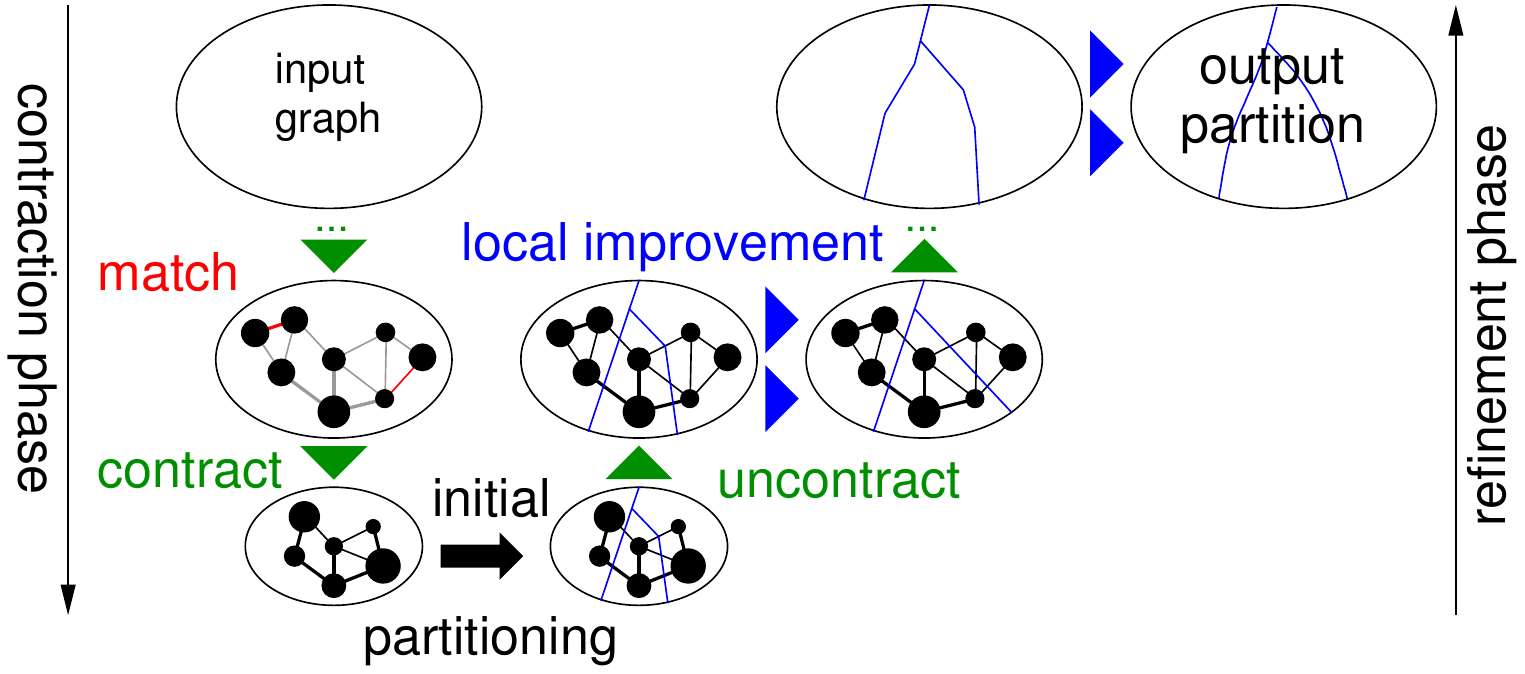}
\end{center}
\label{fig:mgp}
\vspace*{-.5cm}
\caption{Multilevel graph partitioning.}
\vspace*{-.5cm}
%\end{figure}
\end{wrapfigure}

A successful heuristic for partitioning large graphs is the \emph{multilevel graph partitioning} (MGP) approach depicted in Figure~\ref{fig:mgp},
where the graph is recursively \emph{contracted} to 
obtain smaller graphs which should reflect the same basic structure as the input graph. After applying an \emph{initial partitioning} algorithm to the smallest graph, the contraction is undone and, at each level, a
\emph{local search} method is used to improve the partitioning induced by the coarser level.

Recently the partitioning of \emph{complex networks}, such as social
networks or web graphs, has become a focus of
investigation~\cite{costa2011analyzing}. While partitioning meshes is a 
mature field, the structure of complex networks poses new challenges.
Complex networks are often \emph{scale-free} (many low-degree
nodes, few high-degree nodes) and have the {small-world property}.
Small world means that the network has a small
diameter, so that the whole graph is discovered within a few hops from any source node.
These two properties distinguish complex networks from traditional meshes and make finding
small cuts difficult.
Yet, to cope with massive network data sets in reasonable time, there is a need 
for parallel algorithms. Their efficient execution requires good graph partitions.

The paper is organized as follows.
We begin in Section~\ref{s:preliminaries} by introducing basic concepts and by summarizing related work. % in Section~\ref{s:related}.
The main parts of the paper are Sections~\ref{s:sequentialcontraction} and~\ref{s:algorithmicaugmentations}. 
The former introduces our rationale and the resulting size-constrained label propagation algorithm. 
We employ this algorithm during \emph{both} coarsening \emph{and} uncoarsening. 
During coarsening we iteratively compute a size-constrained graph clustering and contract it, and during uncoarsening the algorithm can be used as a fast local search algorithm. 
Section~\ref{s:algorithmicaugmentations} augments the basic algorithm by several algorithmic components. 
The presented algorithms speed up computations and improve solution quality, in particular on graphs that have a irregular structure such as social networks or web graphs. 
Experiments in Section~\ref{s:experiments} indicate that our algorithms are able provide excellent partitioning quality in a short amount of time.
For example, a web graph with 3.3 billion edges can be partitioned on a single machine in about ten minutes while cutting less than half of the edges than the partitions computed by kMetis.
Finally, we conclude with Section~\ref{s:conclusion}.

%%%%%%%%%%%%%%%%%%%%%%%%%%%%%%%%%%%%%%%%%%%%%%%%%
\section{Preliminaries}\label{s:preliminaries}
\vspace*{-.25cm}
\subsection{Basic concepts}
Consider an undirected graph $G=(V=\{0,\ldots, n-1\},E,c,\omega)$ 
with edge weights $\omega: E \to \MdR_{>0}$, node weights
$c: V \to \MdR_{\geq 0}$, $n = |V|$, and $m = |E|$.
We extend $c$ and $\omega$ to sets, \ie 
$c(V')\Is \sum_{v\in V'}c(v)$ and $\omega(E')\Is \sum_{e\in E'}\omega(e)$.
$\Gamma(v)\Is \setGilt{u}{\set{v,u}\in E}$ denotes the neighbors of $v$.
We are looking for \emph{blocks} of nodes $V_1$,\ldots,$V_k$ 
that partition $V$, i.e., $V_1\cup\cdots\cup V_k=V$ and $V_i\cap V_j=\emptyset$
for $i\neq j$. The \emph{balancing constraint} demands that 
$\forall i\in \{1..k\}\gilt c(V_i)\leq L_{\max}\Is (1+\epsilon)c(V)/k+\max_{v\in V} c(v)$ for
some parameter $\epsilon$. 
The last term in this equation arises because each node is atomic and therefore a deviation of the heaviest node has to be allowed.
Note that for unweighted graphs the balance constraint becomes 
$\forall i\in \{1..k\}\gilt |V_i| \leq (1+\epsilon)\lceil\frac{|V|}{k}\rceil$.
The objective is to minimize the total \emph{cut} $\sum_{i<j}w(E_{ij})$ where 
$E_{ij}\Is\setGilt{\set{u,v}\in E}{u\in V_i,v\in V_j}$. 
We say that a block $V_i$ is \emph{underloaded} if $|V_i| < L_{\max}$ and \emph{overloaded} if $|V_i| > L_{\max}$.
A clustering is also a partition, however, $k$ is usually not given in advance and the balance constraint is removed.
A size-constrained clustering constrains the size of the blocks of a clustering by a given upper bound $U$ such that $c(V_i) \leq U$. 
Note that by adjusting the upper bound one can somehow control the number of blocks of a feasible clustering. 
For example, when using $U=1$, the only feasible size-constrained clustering in an unweighted graphs is the clustering where each node forms a block of its own. 
A node $v \in V_i$ that has a neighbor $w \in V_j, i\neq j$, is a boundary node. 
A \emph{matching} $M\subseteq E$ is a set of edges that do not share any common nodes, \ie the graph $(V,M)$ has maximum degree one.
%\emph{Contracting} an edge $\set{u,v}$ means to replace the nodes $u$ and $v$ by a new node $x$ connected
%to the former neighbors of $u$ and $v$. 
%We set $c(x)=c(u)+c(v)$ so that the weight of a node at each level is the number of nodes it is representing in the original graph. 
%If replacing edges of the form $\set{u,w}$, $\set{v,w}$ would generate two parallel edges $\set{x,w}$, a single edge with
%$\omega(\set{x,w})=\omega(\set{u,w})+\omega(\set{v,w})$ is inserted.
%\emph{Uncontracting} an edge $e$ undoes its contraction. 
By default, our initial inputs will have unit edge and node weights. 
However, even those will be translated into weighted problems in the course of the multilevel algorithm.
In order to avoid tedious notation, $G$ will denote the current state of the graph before and after an (un)contraction in the multilevel scheme throughout this paper.
%An abstract view of the partitioned graph is the so called \emph{quotient graph}, where nodes represent blocks and edges stand for connectivity between blocks. 
%The \emph{weighted} version of the quotient graph has node weights which are set to the weight of the corresponding block and edge weights which are equal to the weight of the edges that run between the respective blocks. 

%\vfill
%\pagebreak

%%%%%%%%%%%%%%%%%%%%%%%%%%%%%%%%%%%%%%%%%%%%%
\subsection{Related Work}
\label{s:related}
There has been a \emph{huge} amount of research on GP so that we refer the reader to \cite{GPOverviewBook,SPPGPOverviewPaper} for most of the material. 
Here, we focus on issues closely related to our main contributions. 
All general-purpose methods that work well on large 
real-world graphs are based on the multilevel principle. 
The basic idea can be traced back to multigrid
solvers for systems of linear equations.
% \cite{Sou35},
Recent practical methods are mostly based on graph theoretic aspects, in
particular edge contraction and local search.  
There are different ways to create graph hierarchies such as matching-based schemes \cite{walshaw2000mpm,karypis1998fast,diekmann2000shape,Scotch} or variations thereof~\cite{Karypis06} and techniques similar to algebraic multigrid \cite{meyerhenke2006accelerating,ChevalierS09,SafroSS12}. 
%We refer the interested reader to the respective papers for more details.
Well-known MGP software packages include Jostle~\cite{walshaw2000mpm}, Metis~\cite{karypis1998fast}, 
 and Scotch~\cite{Scotch}. 

Graph clustering with the label propagation algorithm (LPA) has originally been described by Raghavan
\etal \cite{labelpropagationclustering}, but several variants of the algorithm
exist (e.g.~\cite{GRS06}).
%Recently Ovelg\"onne~\cite{Ovelgonne:2012ly} presented a distributed \emph{Hadoop} implementation of
%ensemble graph clustering using LP as a base
%algorithm~\cite{Ovelgonne:2012ly}.
%This implementation processes a 3.3 billion edge web graph
%in a few hours on a 50 machine Hadoop cluster.
%Moreover, 
In addition to its use for fast graph clustering, LPA has been used to partition networks, \eg by Uganer and Backstrom~\cite{UganderB13}. The authors use partitions obtained by geographic initializations and improve the partition by combining LPA with linear programming.
%Another distributed algorithm for balanced graph partitioning has been proposed by Rahimian \etal \cite{jabeja}.  
%The authors use random initializations as starting point for local search based on a slightly extended node swapping approach.
%However, if the initialization is not balanced, the final partition computed by the algorithm will also be imbalanced.
%%we are not distributed anymore

\hmey{What about Noack-Rotta algorithms? Similar in spirit with cluster-based
contraction and local search refinement.}

\paragraph{KaHIP.}
\label{s:kaHIP}
KaHIP -- Karlsruhe High Quality Partitioning -- is a family of GP programs that tackle the balanced GP problem~\cite{kaHIPHomePage,kabapeE}.  
%We integrate the techniques described in this paper into KaHIP~  to evaluate our new algorithms. Hence, we briefly outline its main components.
%KaHIP implements different algorithms. 
It includes KaFFPa (Karlsruhe Fast Flow Partitioner), which is a matching-based multilevel graph partitioning framework that uses for example flow-based methods and more-localized local searches to compute high quality partitions. 
We integrate our new techniques described in this paper into this multilevel algorithm.
%KaHIP also includes various evolutionary algorithms.
KaHIP also includes KaFFPaE (KaFFPa Evolutionary), which is a parallel evolutionary algorithm and KaBaPE (Karlsruhe Balanced Partitioner Evolutionary), which extends the evolutionary algorithm. 
%Moreover, specialized techniques to partition road networks (Buffoon) are included.
%The algorithms in KaHIP have been able to improve the best known partitioning results in the Walshaw Benchmark~\cite{soper2004combined} for many inputs using a short amount of time to create the partitions.

\vfill

\section{Cluster Contraction}
\label{s:sequentialcontraction}
We are now ready to explain the basic idea of our new approach for creating graph hierarchies, which is targeted at complex network such as social networks and web graphs.
%in a multilevel framework. 
%As we borrow the label propagation concept from complex network analysis, our technique is in particular well-suited for partitioning large social networks or large web-graphs. 
We start by introducing the size-constrained label propagation algorithm, which is used to compute clusterings of the graph.
To compute a graph hierarchy, the clustering is contracted by replacing each cluster with a single node, and the process is repeated recursively until the graph is small.
The hierarchy is then used by our partitioner.
KaHIP uses its own initial partitioning and local search algorithms to partition the coarsest graph and to perform local improvement on each level, respectively.
Due to the way the contraction is defined, it is ensured that a partition of a coarse graph corresponds to a partition of the input network with the same objective and balance. 
Note that cluster contraction is an aggressive coarsening strategy. In contrast to previous approaches, it enables us to drastically shrink the size of irregular networks.
The intuition behind this technique is that a clustering of the graph (one hopes) contains many edges running inside the clusters and only a few edges running between clusters, which is favorable for the edge cut objective.
%, the contracted graph will have good properties for being partitioned. 
Regarding complexity our experiments in Section~\ref{s:experiments} indicate that the number edges per node of the contracted graphs is smaller than the number of edges per node of the input network, and the clustering algorithm we use is fast.

\subsection{Label Propagation with Size Constraints}
\begin{wrapfigure}{r}{0.47\textwidth}
\vspace*{-.5cm}
\centering
\includegraphics[width=.4\textwidth]{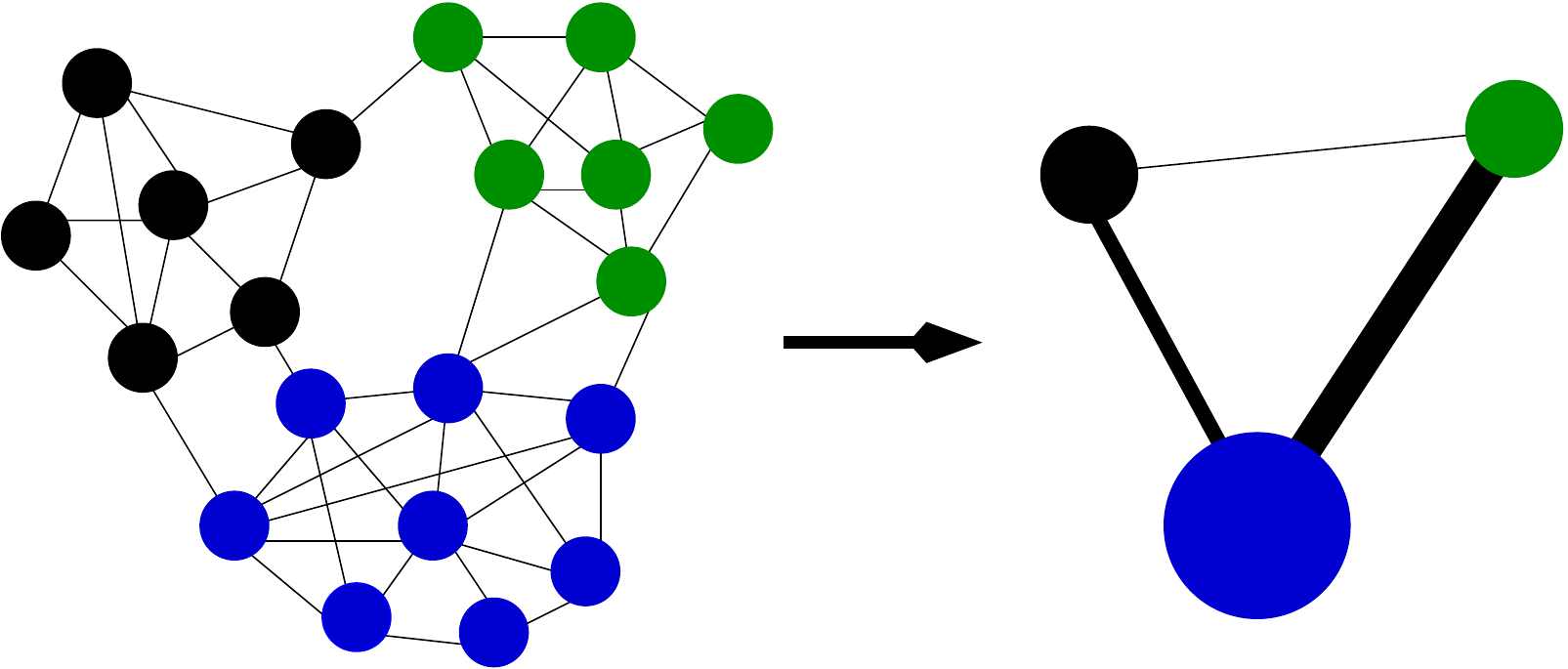}
\caption{Contraction of a clustering. The clustering of the graph on the left hand side is indicated by the colors. Each cluster of the graph  on the left hand side corresponds to a node in the graph on the right hand side.}
\label{fig:clustercontraction}
\vspace*{-.75cm}
\end{wrapfigure}
The \emph{label propagation algorithm} (LPA) was proposed by Raghavan \etal \cite{labelpropagationclustering} for graph clustering. 
It is a fast, near-linear time algorithm that locally optimizes the number of edges cut. We outline the algorithm briefly.  
Initially, each node is in its own cluster/block, \ie the initial block ID of a node is set to its node ID.
The algorithm then works in rounds. 
In each round, the nodes of the graph are traversed in a random order. 
When a node $v$ is visited, it is \emph{moved} to the block that has the strongest connection to $v$, \ie it is moved to the cluster $V_i$ that maximizes $\omega(\{(v, u) \mid u \in N(v) \cap V_i \})$. 
Ties are broken randomly. 
The process is repeated until the process has converged. 
Here, we perform at most $\ell$ iterations of the algorithm, where $\ell$ is a tuning parameter, and stop the algorithm if less then five percent of the nodes changed its cluster during one round.
One LPA round can be implemented to run in $\Oh{n+m}$ time.

After we have computed a clustering, we \emph{contract it} to obtain a coarser graph. 
Contracting the clustering works as follows: 
each block of the clustering is contracted into a single node. 
The weight of the node is set to the sum of the weight of all nodes in the original block. 
There is an edge between two nodes $u$ and $v$ in the contracted graph if the
two corresponding blocks in the clustering are adjacent to each other in $G$,
\ie block $u$ and block $v$ are connected by at least one edge.
The weight of an edge $(A,B)$ is set to the sum of the weight of edges that run between block $A$ and block $B$ of the clustering. 
Note that due to the way the contraction is defined, a partition of the coarse graph corresponds to a partition of the finer graph with the same cut and balance. An example is shown in Figure~\ref{fig:clustercontraction}.

In contrast to the original LPA~\cite{labelpropagationclustering}, we have to ensure that each block of the cluster fulfills a size constraint. There are two reason for this.
First, consider a clustering of the graph in which the weight of a block would exceed $(1+\epsilon) \lceil \frac{|V|}{k} \rceil$. 
After contracting this clustering, it would be impossible to find a partition of the contracted graph that fulfills the balance constraint.
Secondly, it has been shown that using more balanced graph hierarchies is beneficial when computing high quality graph partitions~\cite{kappa}.
To ensure that blocks of the clustering do not become too large, we introduce an upper bound $U := \max( \max_v c(v), W)$ for the size of the blocks. 
Here, $W$ is a parameter that will be chosen later. When the algorithm starts to compute a
  graph clustering on the input graph, the constraint is fulfilled since each of the blocks contains exactly one node. 
A neighboring block $V_\ell$ of a node $v$ is called \emph{eligible} if $V_\ell$ will not become overloaded once $v$ is moved to $V_\ell$.
Now when we visit a node $v$, we move it to the \emph{eligible block} that has the strongest connection to $v$. 
Hence, after moving a node, the size of each block is still smaller than or equal to $U$.
Moreover, after contracting the clustering, the weight of each node is smaller or equal to $U$.
One round of the modified version of the algorithm can still run in linear time by using an array of size $|V|$ to store the block sizes.
We set the parameter $W$ to $\frac{L_{\text{max}}}{f}$, where $f$ is a tuning parameter.
%\begin{figure}[b!]
%\end{figure}

We repeat the process of computing a size-constrained clustering and contracting it, recursively. 
As soon as the graph is small enough, \ie the number of remaining nodes is smaller than $\max{(60k, n/(60k))}$, it is initially partitioned by the initial partitioning algorithms provided in KaHIP.
That means  each node of the coarsest graph is assigned to a block. 
KaHIP uses a multilevel recursive bisection algorithm to create an initial partitioning \cite{dissSchulz}. 
Afterwards, the solution is transferred to the next finer level. 
We assign a node of the finer graph to the block of its coarse representative. 
Local improvement methods of KaHIP then try to improve the solution on the current level. 
%\vfill
%\pagebreak

By using a different size-constraint -- the constraint $W := L_{\text{max}}$ of the original partitioning problem  -- the LPA can also be used as a simple and fast local search algorithm to improve a solution on the current level. 
However, one has to make small modifications to handle overloaded blocks. 
%For example, if the initial partition is imbalanced, the default label propagation algorithm may not be able to recover balance. 
We modify the block selection rule when we use the algorithm as local search algorithm in case that 
the current node $v$ under consideration is from an overloaded block $V_\ell$.
In this case it is \emph{moved} to the eligible block that has the strongest connection to $v$ without considering the block $V_\ell$ it is contained in.
%, \ie it is moved to the block $V_i$ that maximizes $\omega(\{(v, u) \mid u \in N(v) \cap V_i, i \neq \ell\})$. 
This way it is ensured that the move improves the balance of the partition (at the cost of the number of edges cut).
%By default, we use the active nodes algorithm from Section~\ref{s:algorithmicaugmentations} when we use it as a local search algorithm.
Experiments in Section~\ref{s:experiments} show that the algorithm is a fast alternative to the local search algorithms provided by KaFFPa.
Moreover, let us emphasize that the algorithm has a large potential to be efficiently parallelized.

\section{Algorithmic Extensions}
\label{s:algorithmicaugmentations}
In this section we present numerous algorithmic extensions to the approach presented above. This includes using different orderings for size-constrained LPA, combining multiple clusterings into one clustering, advanced multilevel schemes, allowing additional amounts of imbalance on coarse levels of the multilevel hierarchy and a method to improve the speed of the algorithm.
\vspace*{-.25cm}
\paragraph*{Node Ordering for Label Propagation.}
The LPA traverses the nodes in a random order and moves a node to a cluster with the strongest connection in its neighborhood to compute a clustering. 
Instead of using a random order, one can use the ordering induced by the node degree (increasing). 
That means that in the first round of the label propagation algorithm, nodes with small node degree can change their cluster before nodes with a large node degree. 
Intuitively, this ensures that there is already a meaningful cluster structure when the LPA chooses the cluster of a high degree node. 
Hence, the algorithm is likely to compute better clusterings of the graph by using node orderings based on node degree.
We also tried other node orderings such as weighted node degree.  The overall solution quality and running time are comparable so that we omit more sophisticated orderings here.

\vspace*{-.3cm}
\paragraph*{Ensemble Clusterings.} 
In machine learning, ensemble methods combine multiple weak classification (or clustering) algorithms to obtain a strong algorithm for classification (or clustering).
Such an ensemble approach has been successfully applied to graph clustering % LPA as base classifier. This means that 
by combining several base clusterings from different LPA runs. These base clusterings are used to decide whether pairs of nodes should belong to the same cluster~\cite{OvelgoenneG13ensemble,staudtmeyerhenke13high}. 
We follow the idea to get \emph{better} clusterings for the coarsening phase of our multilevel algorithm.

Given a number of clusterings, the \emph{overlay clustering} is a clustering in which two nodes belong to the same cluster if and only if they belong to the same cluster in each of the input clusterings. 
Intuitively, if all of the input clusters agree that two nodes belong to the same block, then they are put into the same block in the overlay clustering. On the other hand, if there is one input clustering that puts the nodes into different blocks, then they are put into different blocks in the overlay clustering.
More formally, given clusterings $\{\mathcal{C}_1, \ldots, \mathcal{C}_\ell\}$, we define the overlay clustering as the clustering where each block corresponds to a connected component of the graph $G_\mathcal{E} = (V,E\backslash \mathcal{E})$, where $\mathcal{E}$ is the union of the cut edges of each of the clusterings $\mathcal{C}_i$, \ie all edges that run between blocks in the clusterings $\mathcal{C}_i$.
Related definitions are possible, \eg a cluster does not have to be a connected component.
%Figure~\ref{fig:exampleoverlayclustering} shows an example overlay clustering.
%\begin{figure}[t!]
%\begin{center}
%\includegraphics[width=10cm]{pics/overlay_clustering.pdf}
%\end{center}
%\vspace*{-.25cm}
%\caption{Two clusterings (top) of a graph are combined to one overlay clustering (bottom).
%Two nodes belong to the same cluster in the overlay clustering if and only if they belong to the same cluster in each of input clusterings.  
%}
%\vspace*{-.25cm}
%\label{fig:exampleoverlayclustering}
%\end{figure}
In our ensemble approach we use the clusterings obtained by size-constrained LPA as input to compute the overlay clustering. 
It is easy to see that the number of clusters in the overlay clustering cannot
decrease compared to the number of clusters in each of the input clusterings. 
Moreover, the overlay clustering is feasible w.r.t. to the size constraint if each of the input clusterings is feasible.

Given $\ell$ clusterings $\{\mathcal{C}_1, \ldots, \mathcal{C}_\ell\}$, we use the following approach to \emph{compute} the overlay clustering iteratively. 
Initially, the overlay clustering $O$ is set to the clustering $\mathcal{C}_1$. 
We then iterate through the remaining clusterings and incrementally update the current solution $O$. 
This is done by computing the overlay $\mathcal{O}$ with the current clustering $\mathcal{C}$ under consideration. 
More precisely, we use pairs of cluster IDs $(i,j)$ as a key in a hash map $\mathcal{H}$, where $i$ is a cluster ID of $\mathcal{O}$ and $j$ is a cluster ID of the current clustering $\mathcal{C}$. 
We then iterate through the nodes and initialize a counter $c$ to zero. 
Let $v$ be the current node. 
If the pair $(\mathcal{O}[v],\mathcal{C}[v])$ is not contained in $\mathcal{H}$, we set $\mathcal{H}(\mathcal{O}[v],\mathcal{C}[v])$ to $c$ and increment $c$ by one.
Afterwards, we update the cluster ID of $v$ in $\mathcal{O}$ to $\mathcal{H}(\mathcal{O}[v],\mathcal{C}[v])$.  
Note that at the end of the algorithm, $c$ is equal to the number of clusters contained in the overlay clustering.
Moreover, it is possible to compute the overlay clustering directly by hashing $\ell$-tuples~\cite{staudtmeyerhenke13high}. 
However, we choose the simpler approach here since the computation of a clustering itself already takes near-linear time.

\vspace*{-.25cm}
\paragraph*{Iterated Multilevel Algorithms.} 
A common approach to obtain high quality partitions is to use a multilevel algorithm multiple times using different random seeds 
%for initialization of the coarsening and local search algorithms. 
%One can then simply 
and use the best partition that has been found.
However, one can do better by transferring the solution of the previous multilevel iteration down the hierarchy. 
In the GP context, the notion of V-cycles has been introduced  by Walshaw \cite{walshaw2004multilevel} and later has been augmented to more complex 
cycles~\cite{kaffpa}. 
These previous works use matching-based coarsening and cut edges are not eligible to be matched (and hence are not contracted).
Thus, a given partition on the finest level can be used as initial partition of the coarsest graph (having the same balance and cut as the partition of the finest graph).  
%We also adopt these more advanced multilevel schemes to improve partitioning quality by modifying the size-constraint label propagation algorithm.
For simplicity, we focus on iterated V-cycles which are illustrated in Figure~\ref{fig:iteratedmultilevel}.
%We shortly outline the general approach and the necessary modifications to our algorithm. 
%This ensures non-decreasing partition quality if the local search algorithm guarantees that solution quality never decreases.
%Note that due to the randomization of the coarsening, the hierarchies created in later cycles are usually different, which introduces further diversification for local search.
We \emph{adopt this technique} also for our new coarsening scheme by ensuring that cut edges are not contracted after the first multilevel iteration.
We present more details in Appendix~\ref{s:apdxclustercontractionvcycles}.

%\vfill
% \pagebreak

\vspace*{-.25cm}
\paragraph*{Allowing Larger Imbalances on Coarse Levels.} It is well-known that temporarily allowing larger imbalance is useful to create good partitions \cite{walshaw2000mpm,kabapeE}. 
Allowing an additional amount of imbalance $\hat \epsilon$ means that the balance constraint is relaxed to $(1+\epsilon + \hat \epsilon)\lceil\frac{|V|}{k}\rceil$, where $\epsilon$ is the original imbalance parameter and $\hat \epsilon$ is a parameter that has to be set appropriately.
%Walshaw and Cross \cite{walshaw2000mpm} estimate the amount of additional allowed imbalance on each level of the multilevel algorithm by taking the number of boundary nodes of perfectly partitioned 2D and 3D meshes into account.
We adopt a simplified approach in this context and decrease the amount of additional allowed imbalance level-wise.
In other words the largest amount of additional imbalance is allowed on the coarsest level and it is decreased level-wise until no additional amount of imbalance is allowed on the finest level.  
To be more precise, let the levels of the hierarchy be numbered in increasing order $G_1, \ldots, G_q$ where $G_1$ is the input graph $G$ and $G_q$ is the coarsest graph. 
The amount of allowed imbalance on a coarse level $\ell$ is set to $\hat \epsilon_\ell = \delta / (q-\ell+1)$, where $\delta$ is a tuning parameter.
No additional amount of imbalance is allowed on the finest level.
Moreover, we only allow a larger amount of imbalance during the first V-cycle.

\vspace*{-.2cm}
\paragraph*{Active Nodes.}
The LPA looks at every node in each round of the algorithm.
Assume for now that LPA is run without a size-constraint.
After the first round of the algorithm, a node can only change its cluster if one or more of its neighbors changed its cluster in the previous round (for the sake of the argument we assume that ties are broken in exactly the same way as in the previous round). 
The active nodes approach keeps track of nodes that can potentially change their cluster. 
A node is called \emph{active} if at least one of its neighbors changed its cluster in the \emph{previous round}.
In the first round all nodes are active.
The original LPA is then modified so that only active nodes are considered for movement. 
This algorithm is always used when the label propagation algorithm is used as a local search algorithm during uncoarsening.
A round of the modified algorithm can be implemented with running time linear in the amount of edges incident to the number of active nodes (see Appendix~\ref{s:apdactivenodesrunningtime}). 
\section{Experiments}
\label{s:experiments}
\paragraph*{Methodology.}
We have implemented the algorithm described above using C++. 
%
%Overall, our the new algorithms consists of about 1000 lines of code (not %including the source of KaHIP). 
We compiled it using g++ 4.8.2.  
The multilevel partitioning framework KaFFPa has different configurations. In this work, we look at the Strong and the Eco configuration of KaFFPa. The aim of KaFFPaEco is to be fairly fast and to compute partitions of high quality, whereas KaFFPaStrong targets very high solution quality.  
%To evaluate interactions and relative importance of our algorithmic improvements, we
%start with KaFFPaEco and enable the algorithmic components described in this paper step-by-step.
%When we start with KaFFPaStrong, we only look at the final algorithm which includes all algorithmic components to save running time and KaFFPaStrong itself. 
Unless otherwise mentioned, we perform ten repetitions for each configuration of the algorithm and report the arithmetic average of computed cut size, running time and the best cut found. 
When further averaging over multiple instances, we use the geometric mean in order to give every instance a comparable influence on the final score.  
For the number of partitions $k$, we choose the values used in  \cite{walshaw2000mpm}: 2, 4, 8, 16, 32, 64.
Our default value for the allowed imbalance is 3\% since this is one of the values used in \cite{walshaw2000mpm} and the default value in kMetis.
We performed a large amount of experiments to tune the algorithm's parameters. Their description is omitted due to space constraints. We use the following parameters of the algorithm which turned out to work well: the number of maximum label propagation iterations $\ell$ during coarsening and uncoarsening is set to 10, the factor $f$ of the cluster size-constraint is set to $18$, the number of V-cycles is set to three and the number of ensemble clusterings used to compute an ensemble clustering is set to 18 if $k$ is smaller than 16, to 7 if $k$ is 16 or 32 and to 3 if $k$ is larger than 32. 

%\mytodo{describe the rest of the parameters}

\begin{table}[t]
\scriptsize
\centering
\vspace*{-.125cm}
\begin{tabular}{|l|r|r|r!{\vrule width 3pt}l|r|r|r|}
\hline
graph & $n$ & $m$ & Ref. & graph & $n$ & $m$ & Ref. \\
\hline
\multicolumn{8}{|c|}{Large Graphs}\\
                \hline
p2p-Gnutella04       & \numprint{6405}   & \numprint{29215}  & \cite{snap}                          & citationCiteseer & \numprint{268495} & $\approx$1.2M     & \cite{benchmarksfornetworksanalysis} \\
wordassociation-2011 & \numprint{10617}  & \numprint{63788}  &  \cite{webgraphWS}                                    & coAuthorsDBLP    & \numprint{299067} & \numprint{977676} & \cite{benchmarksfornetworksanalysis} \\
PGPgiantcompo        & \numprint{10680}  & \numprint{24316}  & \cite{snap}                          & cnr-2000         & \numprint{325557} & $\approx$2.7M     & \cite{benchmarksfornetworksanalysis}\\
email-EuAll          & \numprint{16805}  & \numprint{60260}  & \cite{snap}                          & web-Google       & \numprint{356648} & $\approx$2.1M     & \cite{snap}  \\
as-22july06          & \numprint{22963}  & \numprint{48436}  & \cite{benchmarksfornetworksanalysis}      & coPapersCiteseer & \numprint{434102} & $\approx$16.0M    & \cite{benchmarksfornetworksanalysis}                   \\
soc-Slashdot0902     & \numprint{28550}  & \numprint{379445} & \cite{snap}                          & coPapersDBLP     & \numprint{540486} & $\approx$15.2M    & \cite{benchmarksfornetworksanalysis}                   \\
loc-brightkite       & \numprint{56739}  & \numprint{212945} & \cite{snap}                          & as-skitter       & \numprint{554930} & $\approx$5.8M     & \cite{snap}                                            \\
enron                & \numprint{69244}  & \numprint{254449} &  \cite{webgraphWS}                                    & amazon-2008      & \numprint{735323} & $\approx$3.5M     & \cite{webgraphWS} \\
loc-gowalla          & \numprint{196591} & \numprint{950327} & \cite{snap}                          & eu-2005          & \numprint{862664} & $\approx$16.1M    & \cite{benchmarksfornetworksanalysis}\\
coAuthorsCiteseer    & \numprint{227320} & \numprint{814134} & \cite{benchmarksfornetworksanalysis} & in-2004          & $\approx$1.3M     & $\approx$13.6M    & \cite{benchmarksfornetworksanalysis}\\
wiki-Talk            & \numprint{232314} & $\approx$1.5M     & \cite{snap}                          &                  &                   &                   & \\

\hline
\multicolumn{8}{|c|}{Huge Graphs}\\
\hline
 uk-2002  &  $\approx$18.5M & $\approx$262M &   \cite{webgraphWS}& sk-2005 &$\approx$50.6M & $\approx$1.8G&\cite{webgraphWS}    \\
  arabic-2005 &$\approx$22.7M& $\approx$553M &\cite{webgraphWS}  & uk-2007                                                 & $\approx$106M &  $\approx$3.3G  &  \cite{webgraphWS}  \\
                \hline
\end{tabular}
\vspace*{.2cm}
\caption{Basic properties of the graphs test set.}
 \label{tab:scalefreegraphstable}
\vspace*{-.75cm}
\end{table}
\vspace*{-.25cm}
\paragraph*{Instances.}
We evaluate our algorithms on twenty-five graphs that have been collected from \cite{benchmarksfornetworksanalysis,snap,webgraphWS}. 
This includes a number of  citation and social networks as well as web graphs.  
Table~\ref{tab:scalefreegraphstable} summarizes the basic properties of these graphs. We use the large graph set to evaluate the performance of different algorithms in Section~\ref{s:mainresults} and  compare the performance of the fastest algorithms on the huge graphs in Section~\ref{s:hugegraphs}.
%Moreover, we the following graph families for comparisons.
\vspace*{-.25cm}
\paragraph*{System.} 
We use two machines for our experiments: 
\emph{Machine A} is used for our experimental evaluation in Section~\ref{s:mainresults}. It is equipped with two Intel Xeon E5-2670 Octa-Core processors (Sandy Bridge) which run at a clock speed of 2.6 GHz. 
The machine has 64 GB main memory, 20 MB L3-Cache  and 8x256 KB L2-Cache.
\emph{Machine B} is used for the experiments on the huge networks in Section~\ref{s:hugegraphs}. 
It is equipped with four Intel Xeon E5-4640 Octa-Core processors (Sandy Bridge) running at a clock speed of 2.4 GHz. The machine has 1 TB main memory, 20 MB L3-Cache  and 8x256 KB L2-Cache.

\vspace*{-.25cm}
\subsection{Main Results and Comparison to other Partitioning Packages}
\label{s:expcomparison}
\label{s:mainresults}
In this section we carefully compare our algorithms against other frequently used publicly available tools. 
We compare the average and minimum edge cut values produced by all of these tools on the large graphs from Table~\ref{tab:scalefreegraphstable}, as well as their average running time on these graphs. Experiments have been performed on machine A.
For the comparison we used the $k$-way variant of hMetis 2.0 (p1) \cite{hMetis}, kMetis 5.1 \cite{karypis1998fast} and Scotch 6.0.0 \cite{Scotch} employing the quality option.  
% We used the $k$-way variant of hMetis. However, t
In contrast to our algorithm, hMetis  and Scotch often produce imbalanced partitions. Hence, these tools have a slight advantage in the following comparisons because we do not disqualify imbalanced solutions. 
In case of hMetis the partitions are imbalanced in 105 out of 1260 cases (up to 12\% imbalance) and in case of Scotch the partitions are imbalanced in 218 out of 1260 cases (up to 226\% imbalance).
Note that the latest version of kMetis (5.1) improved the balancing on social networks by integrating a 2-hop matching algorithm that can match nodes if they 
\begin{wraptable}{l}{.4\textwidth}
\scriptsize
\centering
\vspace*{-.5cm}
\begin{tabular}{|l||r|r|r|}
\hline
Algorithm & avg. cut & best cut & $t$ [s] \\
                  \hline
                  \hline

\hline
CEcoR       & \numprint{71814.00} & \numprint{67576.20} & \numprint{10.2} \\
CEco        & \numprint{67222.30} & \numprint{64362.71} & \numprint{8.6} \\
CEcoV       & \numprint{66054.66} & \numprint{63243.32} & \numprint{14.3} \\
CEcoV/B     & \numprint{64584.70} & \numprint{61272.41} & \numprint{15.5} \\
CEcoV/B/E   & \numprint{64724.55} & \numprint{61458.39} & \numprint{46.6} \\
CEcoV/B/E/A & \numprint{65060.78} & \numprint{61762.06} & \numprint{41.9} \\

\hline
CFastR       & \numprint{74414.33} & \numprint{69877.59} & \numprint{4.7} \\
CFast        & \numprint{68839.33} & \numprint{65909.11} & \numprint{3.9} \\
CFastV       & \numprint{67587.23} & \numprint{64713.95} & \numprint{5.7} \\
CFastV/B     & \numprint{70514.36} & \numprint{66783.70} & \numprint{5.8} \\

CFastV/B/E   & \numprint{68977.84} & \numprint{65542.55} & \numprint{28.4} \\
CFastV/B/E/A & \numprint{68942.89} & \numprint{65616.23} & \numprint{24.4} \\
\hline
UFast        & \numprint{69169.78} & \numprint{65965.31} & \numprint{1.5} \\
UFastV       & \numprint{67836.57} & \numprint{64877.33} & \numprint{3.0} \\
UEcoV/B      & \numprint{65212.06} & \numprint{61738.90} & \numprint{11.5} \\

\hline
CStrong       & \numprint{60178.71}  & \numprint{58440.74} & \numprint{422.1} \\
UStrong       & \numprint{59935.70}  & \numprint{58199.17} & \numprint{296.4}\\
\hline
\hline
KaFFPaEco     & \numprint{85920.07}  & \numprint{80577.80} & \numprint{36.2} \\
KaFFPaStrong  & \numprint{63141.24}  & \numprint{60623.96} & \numprint{640.8} \\
\hline
\hline
Scotch        & \numprint{104954.86} & \numprint{97596.38} & \numprint{10.6} \\
kMetis         & \numprint{71977.70}  & \numprint{68434.58} & \numprint{0.4} \\
hMetis        & \numprint{65409.81}  & \numprint{63493.79} & \numprint{107.4} \\
\hline
\end{tabular}
%\vspace*{-.5cm}
\caption{Average cut, best cut and running time results for diverse algorithms on the large graphs. 
%Final score cut values are normalized with the final score of UStrong. 
 Configuration abbreviations: V-cycles (V), add. balance on coarse levels (B), ensemble clusterings (E), active nodes during coarsening (A), random node ordering (R). }
\label{tab:experimentalresultstable}
\vspace*{-.5cm}
\end{wraptable}
share neighbors. 

In addition to the default configurations KaFFPaEco and KaFFPaStrong, we have six base configurations, CEco, CFast, CStrong and UEco, UFast, UStrong.
All of these configurations use the new clustering based coarsening scheme with the degree based node ordering.
The configurations having an Eco in their name use the refinement techniques as used in KaFFPaEco, and the configurations having the word Fast in their name use the label propagation algorithm as local search algorithm instead. KaFFPa implements multilevel recursive bipartitioning as initial partitioning algorithm. The configurations starting with a C use the matching-based approach during initial partitioning and the configurations starting with a U use the clustering based coarsening scheme also during initial partitioning. 
We add additional letters to the base configuration name for each additional algorithmic component that is used.
For example, CEcoV/B/E/A is based on CEco and uses V-cycles (V), additional imbalance  on coarse levels (B), ensemble clusterings (E), and the active node approach (A). 
Moreover, if we use random node ordering (R) instead of degree based node ordering for the LP algorithm, we add the letter R to the base configuration.
CStrong uses additional balance on coarse levels and ensemble clusterings for coarsening. It uses the refinement techniques of KaFFPaStrong. UStrong is the same as CStrong but uses the cluster based partitioning approach for initial partitioning.

\vspace*{-.075cm}
Table~\ref{tab:experimentalresultstable} summarizes the results of our experiments. 
First of all, we observe large running time \emph{and} quality improvements when switching from the matching-based coarsening scheme in KaFFPaEco to the new basic label propagation coarsening scheme (CEcoR vs. KaFFPaEco). In this case, running time is improved by a factor 3.5 and solution quality by roughly 20\%. 
Experiments indicate that already the results obtained by the initial partitioning algorithm on the coarsest graphs are  much better than before.
Additionally enabling the node ordering heuristics yields an extra 8\% improvement in solution quality and 20\% improvement in running time (CEcoR vs CEco and CFastR vs. CFast). 
This is due to the fact that the node ordering heuristic improves the results of the size-constrained LPA by computing clusterings that have less edges between the clusters. 
As a consequence the contracted graphs have a smaller amount of total edge weight, which in turn yields better graph hierarchies for partitioning. 
Performing additional V-cycles and allowing additional imbalance on coarse levels improves solution quality but also increases running time (CEco vs. CEcoV vs. CEcoV/B).
However, allowing additional imbalance does worsen solution quality if the size-constrained LPA is used as a refinement algorithm (CFastV vs. CFastV/B). This is caused by the poor ability of label propagation to balance imbalanced solutions.
Using ensemble clusterings can enhance solution quality (CFastV/B vs. CFastV/B/E), but do not have to (CEcoV/B vs. CEcoV/B/E).
Due to the size-constraints, the clusterings computed by the size-constrained
LPA that uses the active nodes approach have more edges between the clusters than the LPA that does not use this technique.
Hence, the active nodes approach improves running time but also degrades solution quality.
Additional speedups are obtained when the label propagation coarsening is also used during initial partitioning, but sometimes solution quality is worsened slightly. 
For example, we achieve a 2.7 fold speedup when switching from CFast to UFast, and switching from CStrong to UStrong improves solution quality slightly while improving running time by 42\%.

We now compare our algorithms against other partitioning packages.
On average, the configuration UEcoV/B yields comparable quality to hMetis while being an order of magnitude faster. 
When repeating this configuration ten times and taking the best cut, we get an algorithm that has comparable running time to hMetis and improves quality by 6\%.
Our best configuration UStrong, cuts 9\% less edges than hMetis and 20\% less edges than kMetis. In this case, hMetis is a factor 3 faster than UStrong. However, when taking the best cut out of ten repetitions of hMetis and comparing it against the average results of UStrong, we still obtain 6\% improvement. 
Overall, Scotch produces the worst partitioning quality among the competitors.  It cuts 75\% more edges than our best configuration UStrong.
kMetis is about a factor 3.5 faster than our fastest configuration UFast, but also cuts more edges than this configuration. 
%\vfill
%\pagebreak
%\vfill

\vspace*{-.125cm}
\subsection{Huge Web Graphs}
\vspace*{-.025cm}
%\vspace*{-.25cm}
\label{s:hugegraphs}
\begin{wraptable}{r}{0.55\textwidth}
\scriptsize
\centering
\vspace*{-.75cm}
\begin{tabular}{|l||r|r|r||r|r|r|}
\hline
graph           & \multicolumn{3}{c||}{arabic-2005} & \multicolumn{3}{c|}{uk-2002}   \\
\hline
algorithm & avg. cut           & best cut           & $t$ [s] & avg. cut           & best cut     & avg. $t$ [s]     \\
\hline
UFast     & \numprint{1.91}M & \numprint{1.87}M & 111.2   & \numprint{1.47}M & \numprint{1.43}M & 71.7     \\
UFastV    & \numprint{1.85}M & \numprint{1.79}M & 334.3   & \numprint{1.43}M & \numprint{1.39}M & 215.9    \\
\hline
kMetis     & \numprint{3.58}M & \numprint{3.5}M & 99.6    & \numprint{2.46}M & \numprint{2.41}M & 63.7            \\
\hline \hline
&  \multicolumn{3}{c||}{sk-2005} & \multicolumn{3}{c|}{uk-2007} \\
\hline
UFast     & \numprint{23.01}M & \numprint{20.34}M & 387.1   & \numprint{4.34}M  & \numprint{4.10}M  & 626.5  \\
UFastV    & \numprint{19.82}M & \numprint{18.18}M & 1166.4  & \numprint{4.19}M  & \numprint{3.99}M  & 1756.4 \\
\hline                                                                                                                 
kMetis     & \numprint{19.43}M & \numprint{18.56}M & 405.3   & \numprint{11.44}M & \numprint{10.86}M & 827.6  \\
     \hline
\end{tabular}
%\vspace*{.5cm}
\caption{Avg. perf. on huge networks for $k=16$.   }
\vspace*{-.5cm}
\label{tab:hugesocialresultscuts}
\end{wraptable}
In this section our experiments focus on the huge networks from Table~\ref{tab:scalefreegraphstable} (which have up to 3.3 billion undirected edges). To save running time, we focus on the two fast configurations UFast and UFastV, and fix the number of blocks to $k=16$. We speed up UFast and UFastV even more, by only performing three label propagation iterations (instead of ten) during coarsening. Moreover, we also run kMetis and Scotch on these graphs and did not run hMetis due to the large running times on the set of large graphs.
Scotch crashes when trying to partition sk-2005 and uk-2007-05, and did not finish after 24 hours of computations on the other two graphs.
Table~\ref{tab:hugesocialresultscuts} summarizes the results (for more details see Appendix~\ref{apdx:addtionaltabandfig}). 
In three out of four cases our algorithms outperform kMetis in the number of edges cut by large factors  while having a comparable running time. 
Moreover, in every instance 
the best cut produced by the algorithm UFastV is better than the best cut produced by kMetis.
On average, we obtain 74\% improvement over kMetis. 
The
 largest improvements are obtained obtained on the web graph uk-2007. Here, UFast cuts a factor 2.6 less edges than kMetis  and is about 30\% faster.  
The coarse graph after first contraction has already two orders of magnitude less nodes than the input graph and roughly three orders of magnitude less edges.

Note that related work employing label propagation techniques in ensemble graph clustering needs a few hours on a 50 nodes Hadoop cluster to cluster the graph uk-2007 \cite{o13}, whereas for partitioning we currently need only one single machine for about ten minutes. Interestingly, on all graphs (except sk-2005) already the initial partitioning is much better than the final result of kMetis. For example, on average the initial partition of uk-2007 cuts 4.8 million edges, which improves by a factor of 2.4 on kMetis.
%\begin{table}[h]

%2.6
%0.84
%1.67
%1.87

%\vfill 
  %\pagebreak
\vspace*{-.25cm}
\section{Conclusion and Future Work}
\label{s:conclusion}
Current state-of-the-art multilevel graph partitioners have difficulties when
partitioning massive complex networks, at least partially due to ineffective coarsening.
Thus, guided by techniques used in complex network clustering, we have
devised a new scheme for coarsening based on
the contraction of clusters derived
from size-constrained label propagation. Additional algorithmic adaptations, also
for local improvement, further improve running time and/or solution quality.
%
% Key experimental results
The different configurations of our algorithm give users a gradual choice between 
the best quality currently available and very high speed. The quality of the best 
competitor, hMetis, is already reached with an order of magnitude less running time.
The strengths of our techniques particularly unfold
for huge web graphs, where we significantly improve on kMetis in terms
of solution quality (up to $2.6\times$) with a mostly comparable running time.

% Parallelism
Partitioning is a key prerequisite for efficient large-scale parallel graph algorithms.
As huge networks become abundant, there is a need for their parallel analysis.
Thus, as part of future work, we want to exploit the high degree of parallelism exhibited
by label propagation and implement a scalable partitioner for distributed-memory
parallelism.

%The algorithms presented in this work will be made available for download within the KaHIP graph partitioning framework.

{
\tiny
\vspace*{-.25cm}
\bibliographystyle{plain}
\bibliography{phdthesiscs,refs-parco}

\begin{thebibliography}{10}

\bibitem{Karypis06}
A.~Abou-Rjeili and G.~Karypis.
\newblock {Multilevel Algorithms for Partitioning Power-Law Graphs}.
\newblock In {\em Proc. of 20th Int. Parallel and Distributed Processing
  Symp.}, 2006.

\bibitem{benchmarksfornetworksanalysis}
D.~A. Bader, H.~Meyerhenke, P.~Sanders, C.~Schulz, A.~Kappes, and D.~Wagner.
\newblock {Benchmarking for Graph Clustering and Partitioning}.
\newblock In {\em Encyclopedia of Social Network Analysis and Mining}, to
  appear.

\bibitem{GPOverviewBook}
C.~Bichot and P.~Siarry, editors.
\newblock {\em Graph Partitioning}.
\newblock Wiley, 2011.

\bibitem{BuiJ92}
T.~N. Bui and C.~Jones.
\newblock {Finding Good Approximate Vertex and Edge Partitions is {N}{P}-Hard}.
\newblock {\em Information Processing Letters}, 42(3):153--159, 1992.

\bibitem{SPPGPOverviewPaper}
A.~Bulu\c{c}, H.~Meyerhenke, I.~Safro, P.~Sanders, and C.~Schulz.
\newblock {Recent Advances in Graph Partitioning}.
\newblock In {\em Algorithm Engineering -- Selected Topics, \emph{to app.},
  ArXiv:1311.3144}, 2014.

\bibitem{ChevalierS09}
C.~Chevalier and I.~Safro.
\newblock {Comparison of Coarsening Schemes for Multilevel Graph Partitioning}.
\newblock In {\em Proc. of the 3rd Int. Conference on Learning and Intelligent
  Optimization}, volume 5851 of {\em LNCS}, pages 191--205, 2009.

\bibitem{costa2011analyzing}
L.~F. Costa, O.~N Oliveira~Jr, G.~Travieso, F.~A. Rodrigues, P.~Ribeiro
  V.~Boas, L.~Antiqueira, M.~P. Viana, and L.~Enrique C.~Rocha.
\newblock {Analyzing and Modeling Real-World Phenomena with Complex Networks: A
  Survey of Applications}.
\newblock {\em Adv. in Physics}, 60(3):329--412, 2011.

\bibitem{diekmann2000shape}
R.~Diekmann, R.~Preis, F.~Schlimbach, and C.~Walshaw.
\newblock {Shape-optimized Mesh Partitioning and Load Balancing for Parallel
  Adaptive FEM}.
\newblock {\em Par. Computing}, 26(12):1555--1581, 2000.

\bibitem{Garey1974}
M.~R. Garey, D.~S. Johnson, and L.~Stockmeyer.
\newblock {Some Simplified {N}{P}-Complete Problems}.
\newblock In {\em Proc. of the 6th ACM Symp. on Theory of Computing}, STOC '74,
  pages 47--63. ACM, 1974.

\bibitem{GRS06}
J.R. Gilbert, S.~Reinhardt, and V.~Shah.
\newblock {High-performance Graph Algorithms from Parallel Sparse Matrices}.
\newblock March 2006.

\bibitem{HendricksonK00}
B.~Hendrickson and T.~G. Kolda.
\newblock {Graph Partitioning Models for Parallel Computing}.
\newblock {\em Parallel Computing}, 26(12):1519--1534, 2000.

\bibitem{kappa}
M.~Holtgrewe, P.~Sanders, and C.~Schulz.
\newblock {Engineering a Scalable High Quality Graph Partitioner}.
\newblock {\em Proc. of the 24th Int. Parallal and Distributed Processing
  Symp.}, pages 1--12, 2010.

\bibitem{karypis1998fast}
G.~Karypis and V.~Kumar.
\newblock {A Fast and High Quality Multilevel Scheme for Partitioning Irregular
  Graphs}.
\newblock {\em SIAM J. on Scientific Computing}, 20(1):359--392, 1998.

\bibitem{hMetis}
G.~Karypis and V.~Kumar.
\newblock {Multilevel $k$-Way Hypergraph Partitioning}.
\newblock In {\em Proc. of the 36th ACM/IEEE Design Automation Conference},
  pages 343--348. ACM, 1999.

\bibitem{webgraphWS}
University of~Milano Laboratory~of Web~Algorithms.
\newblock Datasets, \url{http://law.dsi.unimi.it/datasets.php}.

\bibitem{snap}
J.~Leskovec.
\newblock Stanford {N}etwork {A}nalysis {P}ackage ({S}{N}{A}{P}).
\newblock \url{http://snap.stanford.edu/index.html}.

\bibitem{meyerhenke2006accelerating}
H.~Meyerhenke, B.~Monien, and S.~Schamberger.
\newblock {Accelerating Shape Optimizing Load Balancing for Parallel FEM
  Simulations by Algebraic Multigrid}.
\newblock In {\em Proc. of 20th Int. Parallel and Distributed Processing
  Symp.}, 2006.

\bibitem{o13}
M.~Ovelg{\"o}nne.
\newblock {Distributed Community Detection in Web-Scale Networks}.
\newblock In {\em 2013 Int. Conf. on Advances in Social Networks Analysis and
  Mining}, pages 66--73, 2013.

\bibitem{OvelgoenneG13ensemble}
M.~Ovelg\"onne and A.~Geyer-Schulz.
\newblock {An Ensemble Learning Strategy for Graph Clustering}.
\newblock In {\em Graph Partitioning and Graph Clustering}, number 588 in
  Contemporary Mathematics. AMS and DIMACS, 2013.

\bibitem{Scotch}
F.~Pellegrini.
\newblock {Scotch Home Page}.
\newblock {\url{http://www. labri.fr/pelegrin/scotch}}.

\bibitem{labelpropagationclustering}
U.~N. Raghavan, R.~Albert, and S.~Kumara.
\newblock {Near Linear Time Algorithm to Detect Community Structures in
  Large-Scale Networks}.
\newblock {\em Physical Review E}, 76(3), 2007.

\bibitem{SafroSS12}
I.~Safro, P.~Sanders, and C.~Schulz.
\newblock {Advanced Coarsening Schemes for Graph Partitioning}.
\newblock In {\em Proc. of the 11th Int. Symp. on Experimental Algorithms
  (SEA'12)}, volume 7276 of {\em LNCS}, pages 369--380. Springer, 2012.

\bibitem{kaHIPHomePage}
P.~Sanders and C.~Schulz.
\newblock {KaHIP -- Karlsruhe High Qualtity Partitioning Homepage}.
\newblock {\url{http://algo2.iti.kit.edu/documents/kahip/index.html}}.

\bibitem{kaffpa}
P.~Sanders and C.~Schulz.
\newblock {Engineering Multilevel Graph Partitioning Algorithms}.
\newblock In {\em Proc. of the 19th Euro. Symp. on Algorithms}, volume 6942 of
  {\em LNCS}, pages 469--480. Springer, 2011.

\bibitem{kabapeE}
P.~Sanders and C.~Schulz.
\newblock {Think Locally, Act Globally: Highly Balanced Graph Partitioning}.
\newblock In {\em Proc. of the 12th Int. Symp. on Experimental Algorithms
  (SEA'12)}, LNCS. Springer, 2013.

\bibitem{dissSchulz}
Christian Schulz.
\newblock {\em {High Quality Graph Partititioning}}.
\newblock PhD thesis, KIT, 2013.

\bibitem{staudtmeyerhenke13high}
C.~L. Staudt and H.~Meyerhenke.
\newblock {Engineering High-Performance Community Detection Heuristics for
  Massive Graphs}.
\newblock In {\em Proc. 42nd Conf. on Parallel Processing (ICPP'13)}, 2013.

\bibitem{UganderB13}
J.~Ugander and L.~Backstrom.
\newblock {Balanced Label Propagation for Partitioning Massive Graphs}.
\newblock In {\em 6'th Int. Conf. on Web Search and Data Mining (WSDM'13)},
  pages 507--516. ACM, 2013.

\bibitem{walshaw2004multilevel}
C.~Walshaw.
\newblock {Multilevel Refinement for Combinatorial Optimisation Problems}.
\newblock {\em Annals of Operations Research}, 131(1):325--372, 2004.

\bibitem{walshaw2000mpm}
C.~Walshaw and M.~Cross.
\newblock {Mesh Partitioning: A Multilevel Balancing and Refinement Algorithm}.
\newblock {\em SIAM J. on Scientific Computing}, 22(1):63--80, 2000.

\end{thebibliography}
}
\vfill
\pagebreak
\begin{appendix}
\section{Additional Figures and Tables}
\label{apdx:addtionaltabandfig}
\begin{figure}[h!]
\begin{center}
\includegraphics[width=9cm]{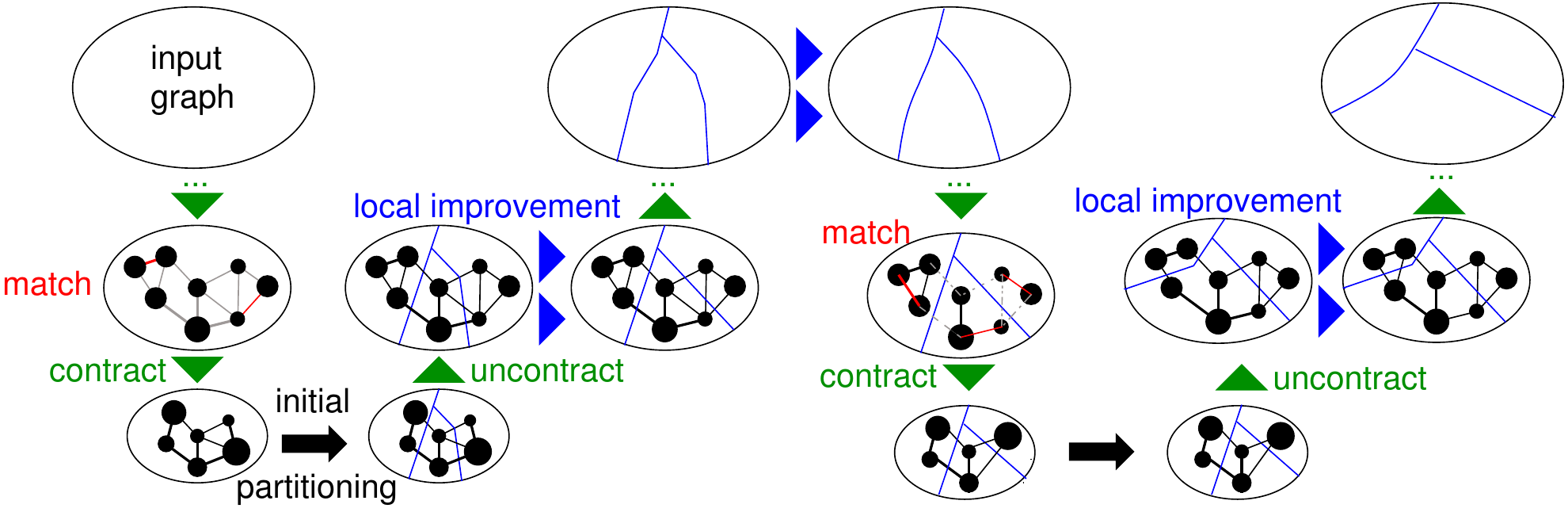}
\end{center}
\caption{Iterated V-cycle Scheme in the matching-based coarsening case. In the second iteration of the multilevel scheme, cut edges are not contracted and the partition computed by the first multilevel iteration can be used as initial partition of the coarsest graph in the second iteration. Thus it is guaranteed that the final partition is at least as good as the partition computed in the first iteration.}
\label{fig:iteratedmultilevel}
\end{figure}
\end{appendix}
\begin{figure}[h!]
\begin{center}
        \includegraphics[width=7cm]{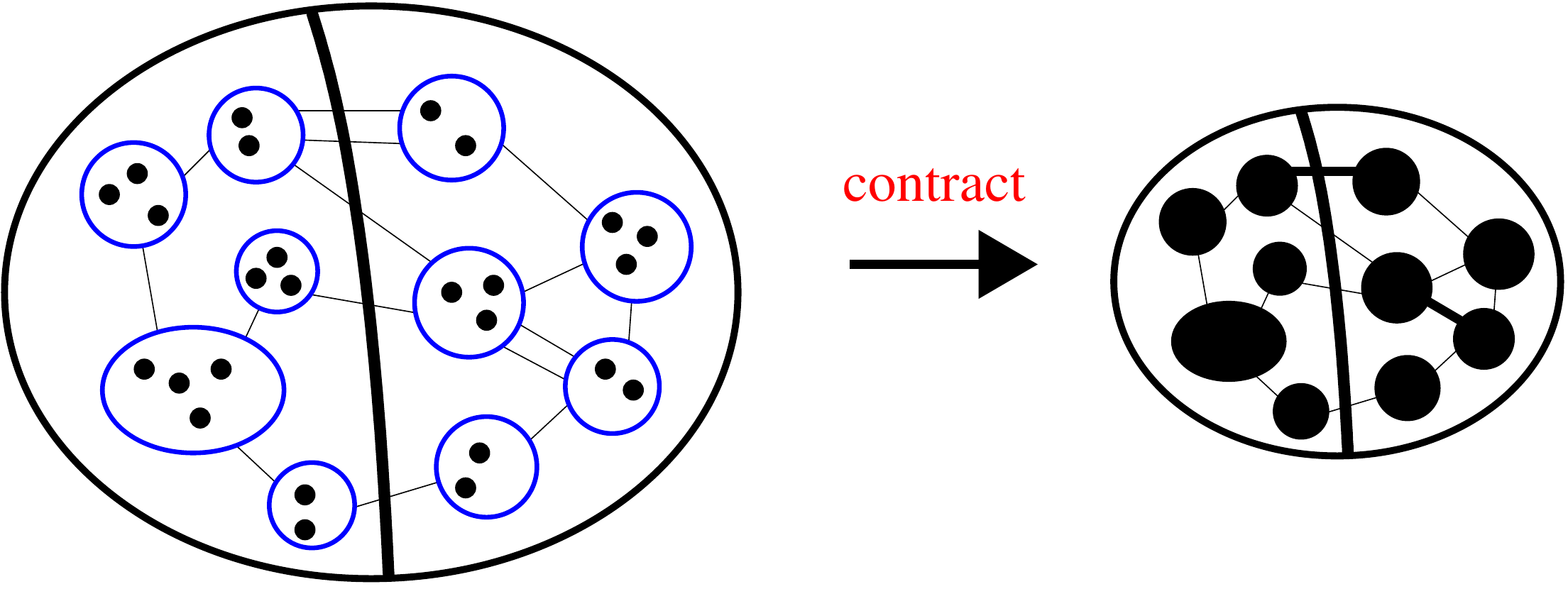}
\end{center}
\caption{A schematic drawing of graph that is partitioned into two blocks (indicated by the thick black line). Each of the clusters (indicated by the blue circles on the left hand side) is a subset of one block of the input partition. Hence, cut edges of the input partition are not contracted, when the clustering is contracted. Moreover, the partition of the graph on the left hand side can be transformed into a partition of the contracted graph that has the same objective and balance.}
\label{fig:respectinputpartition}
\end{figure}
\begin{table}[h!]
\scriptsize
\centering
\vspace*{-.5cm}
\begin{tabular}{|l||r|r|r|r|r|r|}
\hline
                             graph           & \multicolumn{3}{c|}{arabic-2005} & \multicolumn{3}{c|}{uk-2002}   \\
                                     \hline
                             algorithm & avg. cut           & best cut           & avg. $t$ [s] & avg. cut           & best cut     & avg. $t$ [s]     \\
                             \hline
                             %Fast      & \numprint{2087673} & \numprint{1877708} & 266.77   & \numprint{1506629} & \numprint{1481293} & 412.96    \\
                             %FastV     & \numprint{2035709} & \numprint{1845287} & 726.53   & \numprint{1497720} & \numprint{1459560} & 681.37   \\
                             UFast     & \numprint{1914988} & \numprint{1865702} & 111.2   & \numprint{1467201} & \numprint{1432420} & 71.7     \\
                             UFastV    & \numprint{1845030} & \numprint{1790562} & 334.3   & \numprint{1428011} & \numprint{1388692} & 215.9    \\
                             \hline
                             kMetis     & \numprint{3580993} & \numprint{3500691} & 99.6    & \numprint{2459260} & \numprint{2412660} & 63.7            \\
                                     \hline \hline

                             &  \multicolumn{3}{c|}{sk-2005} & \multicolumn{3}{c|}{uk-2007} \\
                                     \hline
%Fast      & \numprint{21279870} & \numprint{18021959} & 1268.41  & \numprint{4378731}  & \numprint{4216993}  & 1247.60  \\
%FastV     & \numprint{19171982} & \numprint{17625346} & 2906.30  & \numprint{4248131}  & \numprint{4079479}  & 6662.77 \\
UFast     & \numprint{23005842} & \numprint{20340871} & 387.1   & \numprint{4341357}  & \numprint{4101463}  & 626.5  \\
UFastV    & \numprint{19818520} & \numprint{18178517} & 1166.4  & \numprint{4190953}  & \numprint{3991890}  & 1756.4 \\
\hline                                                                                                                 
kMetis     & \numprint{19425531} & \numprint{18560644} & 405.3   & \numprint{11441291} & \numprint{10858662} & 827.6  \\

     \hline
\end{tabular}
\vspace*{.5cm}
\caption{Average edge cuts, best cuts and average running times on huge networks for $k=16$.} 
\label{tab:hugesocialresultscutsdetailed}
\end{table}

\vfill
\pagebreak
\section{Additional Technical Details}
\subsection{Adopting V-cycles for Cluster Contraction}
\label{s:apdxclustercontractionvcycles}
To adopt  the iterated multilevel technique for our new coarsening scheme, we have to ensure that cut edges are not contracted after the first multilevel iteration.
We do this by modifying the label propagation algorithm such that each cluster of the computed clustering is a subset of a block of the input partition. In other words, each cluster only contains nodes of one unique block of the input partition. 
Hence, when contracting the clustering, every cut edge of the input partition will remain.
Recall that the label propagation algorithm initially puts each node in its own block so that in the beginning of the algorithm each cluster is a subset of one unique block of the input partition. 
To keep this property during the course of the label propagation algorithm, we restrict the movements of the label propagation algorithm, \ie move a node to an eligible cluster with the strongest connection in its neighborhood that is in the same block of the input partition as the node itself. 
More precisely, let $V=U_1 \cup \ldots \cup U_k$ be the partition of the graph in the current level of the multilevel hierarchy. 
When a node $v \in U_\ell$ is visited, it is moved to an eligible cluster $V_i \subseteq U_\ell$ that maximizes $\omega(\{(v, u) \mid u \in N(v) \cap V_i\})$. Figure~\ref{fig:respectinputpartition} shows a clustering of the graph that respects a given partition.

\subsection{Running Time of the Active Nodes Approach}
\label{s:apdactivenodesrunningtime}
One round of the label propagation algorithm using active nodes algorithm can be implemented with running time linear in the amount of edges incident to the number of active nodes. This is done by using two FIFO queues, one that stores the active nodes of the current round and one for the next round, as well as two bit vectors that store if a node is already contained in the respective queue. 
During one round, the algorithm iterates over all nodes in the FIFO queue and considers them for movement.
Once a node is considered for movement, the entry in the bit vector corresponding to the current queue is set to false.
If the algorithm moves a node, all its neighbors are set active for the next round of the algorithm. 
Moreover, if a neighbor is not already contained in the queue for the next round, it is inserted and the respective bit vector is updated.
When the round is finished, all entries in the bit vector associated with the queue of the current round are false (and the queue is empty). Hence, the queues change its roles, \ie pointers to the queues and the associated bit vectors are swapped.

\end{document}